\documentclass[referee]{raa}
\usepackage{graphicx,times}
\usepackage{natbib}
\usepackage[papersize={8.27in,11.69in}]{geometry}
\usepackage[flushleft]{threeparttable}
\bibpunct{(}{)}{;}{a}{}{,}

\begin{document}

   \title{The variability and radial velocity of planetary nebulae central stars.
}

 \volnopage{ {\bf 20XX} Vol.\ {\bf X} No. {\bf XX}, 000--000}
   \setcounter{page}{1}

   \author{Ali, A.\inst{1}$^*$ \footnotetext{\small $*$ Corresponding author}, Mindil, A.\inst{2}}


\institute{Astronomy, Space Science \& Meteorology Department, Faculty of Science, Cairo University, Giza 12613, Egypt.; {\it afouad@sci.cu.edu.eg}\\
 \and
Department of Physics, College of Science, University of Jeddah, Jeddah, Saudi Arabia.; {\it amindil@uj.edu.sa}\\
\vs \no
   {\small Received 20XX Month Day; accepted 20XX Month Day}
}

\abstract{\\
The extremely accurate estimates of stellar variability and radial velocity in the Gaia Data Release 3 (Gaia DR3) have enabled us to examine the close binarity and radial velocity (RV) of central stars (CSs) of planetary nebulae (PNe). This study is twofold: (1) searching for new close binary CSs candidates to better understand how binarity affects the formation and evolution of PNe; and (2) extending the sample size of known RV of PNe in order to understand their kinematics and the dynamics of the Milky Way. As a target sample, we used all true, possible, and likely PNe available in the literature. Then, we looked for their matched Gaia DR3 sources that provide measurements of variability and RV.   As a result, we detected the first large collection of trustworthy photometric variability of 26 symbiotic stars (SySts) and 82 CSs. In this CS group, there are 24 sources already classified as true close binary CSs in the literature.  Hence, we discovered 58 new close binary CS candidates. This close binary (CB) sample represents more than half of what is currently available in the literature.  In addition, we identified the radial velocities for 51 PNe. To our knowledge, 24 of these were measured for the first time. The RV measurements predicted by Gaia, based on the Doppler shift of the CS absorption lines, and those derived from nebular emission lines, show satisfactory agreement except for a few extremely high-velocity PNe.
}

   \authorrunning{Ali \& Mindil}            
   \titlerunning{variability and radial velocity of PNCSs}  
   \maketitle

\section{Introduction}           

Gaia is an ESA project that aims to create a 3-D representation of the Milky Way galaxy. The Gaia Data Release 1 (Gaia DR1) appeared in September 2016. This was followed by Gaia Data Release 2 (Gaia DR2) in April 2018 and Early Data Release 3 (Gaia EDR3) in December 2020. The Gaia DR3 was published on June 13, 2022. It provides positions and apparent magnitudes of $\sim 1.8$ billion sources, as well as parallaxes, proper motions, and colors of $\sim 1.5$ billion objects. In comparison to Gaia DR2, Gaia DR3 exhibits significant improvements in astrometric and photometric accuracy, precision, and homogeneity. In addition to updating earlier releases, Gaia EDR3 contains new data, such as astrophysical parameters \citep{Creevey22}, BP/RP spectra \citep{De_Angeli22}, and variability classification \citep{Eyer22}. According to \citet{Eyer22}, Gaia DR1 has $\sim 3000$, Gaia DR2 has $\sim 550,000$, and Gaia DR3 has $\sim 10.5$ million variable sources.

The topic of binarity has important implications for our understanding of cataclysmic variables and novae, type Ia supernovae, symbiotic stars, and other phenomena such as the production of astrophysical jets \citep{Boffin19}. Binary interactions occur between stars of all sizes and orbital separations, ranging from compact white dwarfs with 5-minutes orbital periods to giant stars with hundred-years orbital periods. From an observational and theoretical perspective, all stars with masses ranging from 1 to 8 solar masses—roughly 95\% of the galaxy's stellar population—will undergo the PN stage of evolution. As a consequence, studying the binary CSs of PNe is crucial to our understanding of many astrophysical phenomena that have traditionally been attributed to single-star evolution \citep{Aller20}.

The HASH catalog  \citep{Parker16} contains $\sim 3500$ PNe, with 80\% displaying complex morphologies that differ from sphericity, such as elliptical, bipolar, and multipolar PNe, as well as various internal features such as multi-shells, jets, and knots. Currently, there is widespread agreement on the importance of CS binarity in understanding PNe divergence from sphericity, where the various morphologies of PNe can no longer be explained by single-star models. Since the number of detected binary CSs has increased significantly over the past decade, it has become obvious that the wide variety of PNe morphologies and some of their unusual chemical properties are the products of binary evolution. A common-envelope event is the best method for generating an axisymmetric PN via binary interaction. The CB fraction is the most rigorous binarity test in PN formation. Despite the challenge of finding CS infrared excesses, \citet{DeMarco13} successfully employed the technique to calculate a binary fraction, obtaining a value of 67–78\% based on I-band excess and 100–107\% based on J-band excess. Douchin et al. (2015) obtained I-band fraction of 40$\pm$20\% and J-band fraction of 62$\pm$30\% using an improved method and a larger sample. Estimates of the binary fraction range from 20\% for photometrically detectable CB to 60-80\% for those identified using the radial velocity variability and infrared excesses approaches. According to \citet{Boffin19}, the bipolar and multipolar PNe are the result of CB stellar evolution. \citet{Wesson18} identified a link between the high abundance discrepancy factors (adfs) of PNe and their CSs binarity. It was found that all PNe of binary CSs with a period less than 1.15 days had adfs larger than 10 and electron densities of less than 1000 cm$^{-3}$, whereas those with longer periods had lower adfs and significantly higher electron densities. In addition, they noted that any PN with an extreme adf must contain a close binary CS.

In addition to ground-based observations of close binary CSs (e.g. \citet{Hillwig16}, \citet{Miszalski11a}, \citet{Jones10}, \citet{Miszalski09}), space-based observations, such as those from the Transiting Exoplanet Survey Spacecraft-TESS \citep{Aller20} and Kepler satellite \citep{Jacoby21}, have detected a significant number of CB candidates. Most of the photometric variability of these objects can be attributed to the effect of a companion star on the nebular CS.

Gaia DR3 has radial velocities for $\sim 34$ million stars and RV spectrometer spectra for almost a million stars \citep{Katz22}. The Gaia Data Release 4 (Gaia DR4), which will analyze 66 months of data, will extend all RV spectra to a G-magnitude of 16.2 and reveal the RV of $\sim 100$ million stars.

In the present article, we aim to uncover the variability and RV of the CSs of PNe using the recent release of the Gaia project. We showed the PNe data sample and the approach used for extracting the variability and RV from the Gaia DR3 database in Section 2. Section 3 contains the results and discussion, whereas Section 4 has the conclusions.

\section{The variability and RV data.}

The RV spectrometer is a medium-resolution spectrograph (R $\approx 11 500$) covering the wavelength range 846 – 870 nm \citep{Cropper18}.  In total,$\sim 10.5$  million objects have been identified as variables in the Gaia DR3. \citet{Eyer22} have reported the presence of 35 types and sub-types of variable objects, where the output of the variability analysis amounts to 17 tables containing a total of 365 parameters. The stellar photometric variability is stored in the gaiadr3.gaia$\_$source table in the field phot$\_$variable$\_$flag. The combined RVs and their formal uncertainties are, respectively, stored in the radial$\_$velocity and radial$\_$velocity$\_$error fields in the gaiadr3.gaia$\_$source table.

To achieve the goals of this article, we searched the Gaia DR3 database for all stars whose positions matched those of PN central stars listed in the HASH catalog as true, possible, and likely PNe.

\section{Results and discussion}

\subsection{Close binary of PN central stars}

\citet{Chornay21B} reported a list of 58 likely close binary CSs using the photometric data in Gaia DR2. They detected the variability of these objects not as a direct result of extracting the phot$\_$variable$\_$flag identifier in the gaiadr2.gaia$\_$source module, but according to a method that depends on the flux, magnitude, and color uncertainties of the object (for more details, see \citet{Chornay21B}.

Using the database sample, which is composed of roughly 3500 PNe, we found 113 CSs showing photometric variability through the phot$\_$variable$\_$flag identifier in the gaiadr3.gaia$\_$source table. Looking for more information on each PN in the SIMBAD database and the HASH catalog, we noticed that 27 PNe have been re-classified as SySts and four as M type, Hot subdwarf, Mira variable, and Wolf-Rayet stars (see Table 2). The remaining 82 CSs are associated with 75 true, 4 likely, and 3 possible PNe (Table 1). From this list, there are 24 CSs have been documented as possessing CB systems in the literature. As a result, we have detected 58 new close binary CS candidates. This set represents more than half of the known closed binary CSs \citep{Boffin19}.

The binarity of CS may be inferred from its color. The central star is often blue owing to its enormous UV radiation, but there are also a lot of red CSs. This might be explained by the fact that the visible light of the main sequence or red giant companion dominates the CS color.  Table 1 shows that approximately 70\% of the CSs are red (B-R $> 0.0$), implying that they are possibly close binaries. In addition, Table 1 lists the periodicity time and the reference of each CS, which is considered a true close binary in the literature. Furthermore, we examined the list of variable CSs presented by \citet{Chornay21B}, where we found only 4 stars (listed in Table 1, in boldface style) were explicitly defined as variables using the phot$\_$variable$\_$flag identifier.

The morphological type of each PN that was retrieved from the HASH catalog is given in Table 1. As predicted by most current theories, the majority of suspected close binary CSs (85\%) are surrounded by bipolar and elliptical nebulae \citep{Boffin19}. Moreover, $\sim 50\%$ of these nebulae have multiple shells. In addition, Table 1 contains 9 PNe previously identified as having CB central stars and high adfs (Hf\,2-2; A66\,41; A66\,63; K\,1-2; Sp\,1; HaTr\,4, Hen\,2-248; NGC\,6026; M\,3-16).

Symbiotic stars have the longest orbital periods of all interacting binaries. It consists of an evolved, cool star transferring mass to a much hotter, brighter, compact partner \citep{Ilkiewicz17}. Because the spectra of SySts are similar to those of PNe and HII regions, all the objects in Table 2 were previously thought to be PNe. It is worth noting that all SySts in Table 2 are red in color.

\subsection{Radial velocity of PNCSs}

The pioneering work for determining the RV of PNe was given by \citet{Schneider83} who published the heliocentric RV for 524 PNe. The next compilation (867 PNe) was reported by \citet{Durand98}. \citet{Beaulieu99} reported the RV of 45 PNe lying in the southern galactic bulge. Based on high dispersion spectra, \citet{Richer17} reported the RV of 76 PNe. Numerous other individual RVs are scattered throughout the literature  (e.g. \citealt{Ali16,Ali17,Ali19}). All the above measurements were derived from the Doppler shift of the nebular spectra. The Gaia mission opened a new window for calculating the RV from the spectra of the observed CSs.  The Gaia RV was found by measuring the Doppler shift of a template spectrum and then comparing it to the spectrum that was seen.

Using the current release, we were able to detect the RV for 51 PNe, including updated values for 14 PNe recorded by \citet{Ali22}. Table 3 lists the newly detected radial velocities by Gaia DR3 and those obtained by \citet{Durand98}. The estimated median uncertainty of this compilation is 12.2\%. In Figure 1, we compared the new RV measurements with those given by \citet{Durand98}. The diagonal line indicates the 1:1 matches. In general, the RVs computed from the spectra of both the CSs (Gaia DR3) and their associated nebulae are in good agreement. However, there are a few outliers related to high-velocity objects, such as H\,2-24, SB\,15, and Th\,3-14. At first glance, all outlier objects have galactic longitudes of 0 to 10 and 350 to 360 degrees, and galactic latitudes of 0 to $\pm$ 10 degrees, indicating that they are located in the direction of the galactic bulge. To figure out the cause of this discrepancy, we examined the possible physical reasons, such as the interaction between planetary nebulae and the interstellar medium (ISM), the effect of nebular electron density, and the accuracy of radial velocity measurements deduced from PNe and CSs. We found that none of these nebulae interact with the ISM, and the available electron density data for these objects did not provide a reasonable explanation. In addition, the accuracy of RV measurements derived from nebular emission lines is suitable. Thus, we examined the parameters that Gaia used to calculate the RV. We extracted two additional parameters relevant to the Gaia RV calculations: rv$\_$nb$\_$transits (the number of transits used to calculate the RV) and rv$\_$visibility$\_$periods$\_$used (the number of visibility periods used to estimate the RV). Table 3, shows the previous two parameter values in columns 6 and 7. The preceding two parameters show that these outliers have a lower number of transits and shorter visibility periods compared with other RV measurements in Table 3. As a result, we may infer that the difference in RV measurements between the Gaia and nebular lines for these outlier objects is due to Gaia's inaccurate RV measurements.

\section{Conclusions}

We have discovered 82 planetary nebulae associated with close binary central star candidates. To our knowledge, 58  members of this group have been found for the first time. This group of close binary central stars comprises roughly half of all objects known in the literature. We also discovered photometric variability in 26 symbiotic stars and four stars of different types.  Moreover, we detected the radial velocities of 51 planetary nebulae, 27 of which were identified for the first time. With a few exceptions, there is good agreement between the radial velocities measured from the absorption lines of the central stars and those measured from the emission lines of the planetary nebulae.
In future work, we plan to extract the available photometric variability identifiers from the Gaia DR3 database to build the light curves for some of the objects mentioned in Table 1. We also plan to use the 74-inch telescope at the Kottami observatory, Egypt, to perform a time-series photometric study for a few of the detected close binary central star candidates. Simultaneously, we will search the TESS and Kepler Sky Surveys, as well as the OGLE variable star catalog, for data that will allow us to confirm the binarity of the newly detected objects.
\begin{acknowledgements}
The authors would like to thank the reviewer for his or her constructive suggestions that helped enhance the original manuscript. 
\end{acknowledgements}

\bibliographystyle{raa}
\bibliography{bibtex-msRAA-2022-0313}

\begin{figure}
\includegraphics[scale=0.40]{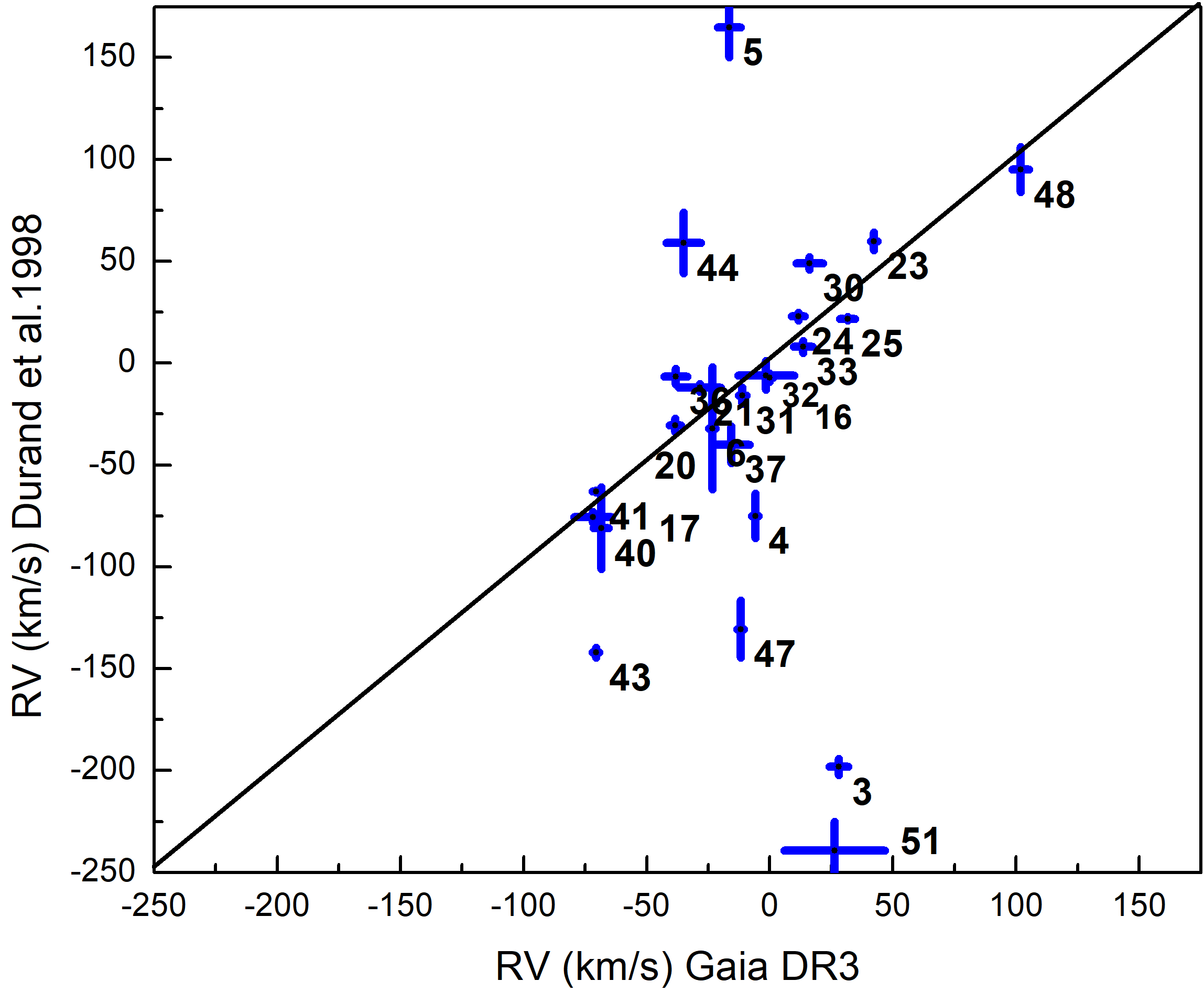}
\caption{The RVs derived from the Gaia DR3 against those reported by \citet{Durand98}. The diagonal line refers to the 1:1 correlation. The numbers on the plot assign the PNe numbers that listed in Table\,3.}
\label{Figure2}
\end{figure}

\begin{table*}
\tiny
\centering
\caption{\small{The variable CSs in Gaia DR3. The magnitude and color of the CS are indicated by the G and B-R parameters, respectively. The letters T, L, and P stand for true, likely, and possible PNe. Bipolar, elliptical, round, irregular, and quasi-Stellar PN shapes are denoted by the primary morphological keys B, E, R, I, and S, respectively. Multiple shells, point symmetry, ring structure, and asymmetry are denoted by the internal structure symbols m, p, r, and s, respectively.}} \label{Table1}
\begin{threeparttable}
\scalebox{0.7}{
\begin{tabular}{llllllllllllllllll}
\hline
PN Name	&	l	&	b	&	 Gaia DR3 designation	&	G	&	B-R	&	PN Status	&	Shape	&	Ref.	& 	PN Name	&	l	&	b	&	 Gaia DR3 designation	&	G	&	B-R	&	PN Status	&	Shape	&	Ref.\\[3pt]
\hline
PN PC 12	&	0.17	&	17.25	&	 4130784921205604736	&	15.2	&	0.6	&	T	&	Bmpr	&	1	&	IC 2165	&	221.32	&	-12.39	&	2999839084924027776	&	17.5	&	-0.2	&	T	&	Emrs	&		\\
PN Bl O	&	0.88	&	-1.57	&	 4056603178196321792	&	16.5	&	1.8	&	T	&	S	&		&	PFP 1	&	222.13	&	3.91	&	 3058094200264637312	&	15.8	&	-0.6	&	T	&	Rar	&		\\
PPA J1800-2904	&	1.52	&	-2.85	&	 4062356711999251328	&	18.1	&		&	T	&	s	&		&	PN M  3-2	&	240.37	&	-7.63	&	 5609860130542365824	&	16.3	&	0.5	&	T	&	Bms	&		\\
PN ShWi 7	&	1.80	&	-3.88	&	 4050366645122261504	&	17.9	&	1.2	&	T	&	B	&	2	&	PG 1034+001	&	247.55	&	47.75	&	 3806885288337214848	&	13.2	&	-0.6	&	T	&	na	&	11	\\
Terz N 2111	&	3.96	&	1.66	&	 4067312696253910272	&	16.9	&	4.4	&	T	&	Ea	&		&	PN K  1-2	&	253.58	&	10.78	&	 5647809392112960000	&	17.0	&	0.2	&	T	&	Baps	&	12	\\
PN H 2-24	&	4.33	&	1.84	&	 4068460105422978048	&	15.3	&	4.3	&	T	&	Ba	&		&	PN M  3-6	&	253.97	&	5.78	&	 5639472001599302528	&	13.2	&	0.0	&	T	&	E	&		\\
PN Hf 2-2	&	5.14	&	-8.90	&	 4048497024309080064	&	17.2	&	0.0	&	T	&	Ems	&	3	&	LoTr 3	&	265.11	&	-4.21	&	 5521499734013833984	&	13.0	&	0.7	&	T	&	Rr	&		\\
PN H 2-22	&	6.34	&	3.33	&	 4117062676912301184	&	18.5	&	1.6	&	T	&	B	&		&	PN Lo 4	&	274.31	&	9.11	&	 5414927915911816704	&	16.6	&	-0.4	&	T	&	Ears	&		\\
PN PBOZ 29	&	6.59	&	3.41	&	 4118615354715439872	&	16.1	&	1.6	&	L	&	S	&		&	Wray 16-55	&	277.62	&	-1.73	&	 5308685822467307008	&	12.2	&	5.9	&	T	&	S	&		\\
NGC 6629	&	9.41	&	-5.05	&	 4089517157442187008	&	12.7	&	0.5	&	T	&	Ems	&		&	PN G281.1-00.4	&	281.18	&	-0.48	&	 5259854002824501248	&	17.7	&	3.3	&	P	&	na	&		\\
PN A66 41	&	9.66	&	10.51	&	 4136835641106850432	&	16.2	&	0.3	&	T	&	Bas	&	4	&	PN K  1-22	&	283.67	&	25.31	&	 5399388964749811456	&	16.7	&	0.4	&	T	&	Ears	&	13	\\
PN Sa 3-111	&	14.27	&	4.21	&	 4147061232357104384	&	17.0	&	3.7	&	T	&	S	&		&	DS 1	&	283.90	&	9.73	&	 5362804330246457344	&	12.1	&	-0.3	&	T	&	Ims	&	2	\\
PN M  1-46	&	16.45	&	-1.98	&	 4103910524954236928	&	12.8	&	0.8	&	T	&	Rmprs	&		&	\textbf{Hen 2-70}	&	293.61	&	1.20	&	 5335879596943573888	&	15.7	&	2.0	&	T	&	Bamps	&		\\
PTB 43	&	16.62	&	-4.05	&	 4102825336944868480	&	17.3	&	0.9	&	T	&	S	&		&	NGC 4361	&	294.11	&	43.63	&	 3519614068578061568	&	13.1	&	-0.5	&	T	&	Emps	&		\\
PN PM 1-308	&	34.58	&	-11.75	&	 4210278482327706496	&	13.1	&	0.6	&	T	&	na	&		&	PN G305.9-01.2	&	305.93	&	-1.27	&	 5859151160662602752	&	19.3	&	2.9	&	T	&	B	&		\\
\textbf{PN G039.0-04.0}	&	39.08	&	-4.10	&	 4292267621344388864	&	14.3	&	1.5	&	T	&	Emr	&		&	Hen 2-99	&	309.00	&	-4.24	&	 5851865148069389568	&	13.2	&	0.4	&	T	&	Ers	&		\\
VSP 2-30	&	49.32	&	2.38	&	 4320639728629291776	&	13.6	&	1.2	&	L	&	S	&		&	PN SuWt 2	&	311.05	&	2.48	&	 5870592987893097984	&	11.9	&	0.6	&	T	&	Eamrs	&		\\
PN A66 63	&	53.89	&	-3.03	&	 1820963913284517504	&	15.0	&	0.3	&	T	&	Bps	&	5,6,7	&	Hen 2-107	&	312.61	&	-1.90	&	 5854138766383247232	&	14.6	&	1.2	&	T	&	Ea	&		\\
NGC 6891	&	54.20	&	-12.11	&	 1803234906762692736	&	12.3	&	-0.2	&	T	&	Emrs	&	8	&	PN Sp 1	&	329.08	&	1.96	&	 5982072132545824128	&	13.7	&	0.7	&	T	&	Ramrs	&	3,4	\\
PN G054.5+01.8	&	54.59	&	1.85	&	 4515887189511585792	&	18.6	&	1.4	&	L	&	E	&		&	\textbf{PN Mz 3}	&	331.73	&	-1.01	&	 5934701559547878144	&	13.2	&	1.8	&	P	&	na	&		\\
PN A66 46	&	55.41	&	16.03	&	 4585381817643702528	&	15.0	&	-0.2	&	T	&	Eas	&	5,6,7	&	PN HaTr 7	&	332.51	&	-16.91	&	 5911656865276078080	&	14.8	&	-0.3	&	T	&	Eas	&	3	\\
PN K 3-51	&	56.83	&	-6.96	&	 1821791540605697152	&	17.2	&	-0.1	&	T	&	R	&		&	IC 4642	&	334.39	&	-9.35	&	 5923374773032038528	&	15.9	&	-0.2	&	T	&	Ems	&		\\
IRAS 19461+2419	&	60.99	&	-0.57	&	 2020643612977496704	&	18.6	&	2.5	&	T	&	S	&		&	PN HaTr 4	&	335.25	&	-3.62	&	 5937103069115240192	&	16.8	&	0.7	&	T	&	B	&		\\
NGC 6720	&	63.17	&	13.98	&	 2090486618786534784	&	15.6	&	-0.5	&	T	&	Emrs	&		&	MPA J1637-4911	&	335.95	&	-1.35	&	 5940883018248096640	&	17.2	&	2.2	&	L	&	S	&		\\
PN Ps 1	&	65.02	&	-27.31	&	 1745948362385436544	&	14.7	&	0.3	&	T	&	s	&		&	Hen 2-248	&	341.51	&	-9.18	&	 5946831685377720576	&	15.4	&	-0.3	&	T	&	S	&	2	\\
ETHOS 1	&	68.10	&	10.99	&	 2050526964622031744	&	17.2	&	-0.1	&	T	&	Bmps	&	9,10	&	NGC 6026	&	341.60	&	13.70	&	 6011169161583903488	&	13.1	&	0.1	&	T	&	Eas	&	2	\\
MWP 1	&	80.36	&	-10.41	&	 1855295171732158080	&	13.0	&	-0.5	&	T	&	Baps	&		&	IC 1266	&	345.24	&	-8.83	&	 5954912374289120896	&	11.3	&	0.0	&	T	&	Rars	&		\\
PN M  1-77	&	89.38	&	-2.27	&	 1971995510535755648	&	11.9	&	1.0	&	T	&	Sm	&		&	PN Tc1	&	345.24	&	-8.83	&	 5954912374289120896	&	11.3	&	0.0	&	T	&	Rars	&		\\
PN K 1-16	&	94.03	&	27.43	&	 2160562927224840576	&	15.0	&	-0.6	&	T	&	B	&		&	IC 4637	&	345.48	&	0.14	&	 5966769881320062208	&	12.5	&	0.6	&	T	&	Eaprs	&		\\
NGC 40	&	120.02	&	9.87	&	 537481007814722688	&	11.5	&	0.3	&	T	&	Bms	&		&	PPA J1747-3435	&	355.33	&	-3.21	&	 4041711044735017856	&	18.8	&	0.8	&	T	&	Es	&	2	\\
PB 9	&	122.72	&	70.36	&	 1531053247144552704	&	18.0	&	0.7	&	T	&	Eams	&		&	PN M  1-27	&	356.53	&	-2.39	&	 4053955824662571648	&	13.9	&	1.4	&	T	&	R	&		\\
PB 4	&	123.11	&	70.19	&	 1531068915184317568	&	17.3	&	0.6	&	T	&	Emrs	&		&	PN M 4-4	&	357.03	&	2.44	&	 4058620300987916160	&	16.5	&	3.8	&	T	&	Ear	&	1	\\
PB 1	&	123.18	&	70.08	&	 1531072827896228352	&	17.9	&	0.3	&	T	&	Ems	&		&	PN Al 2-O	&	358.01	&	-2.74	&	 4043622756199128064	&	16.0	&	2.4	&	T	&	E	&	2	\\
NAME TS 01	&	136.00	&	55.97	&	 846615127231002880	&	18.0	&	-0.3	&	T	&	Es	&		&	PN Al 2-R	&	358.75	&	-2.76	&	 4055678213978728320	&	15.4	&	5.3	&	T	&	B	&		\\
PN HFG 1	&	136.38	&	5.55	&	 468033345145186816	&	14.0	&	0.7	&	T	&	Eamrs	&	2	&	PHR J1752-3116	&	358.77	&	-2.50	&	 4055698280071328640	&	16.1	&	3.4	&	T	&	S	&		\\
\textbf{NGC 1501}	&	144.56	&	6.55	&	 473712872456844544	&	14.2	&	0.6	&	T	&	Ems	&		&	JaSt 65	&	358.99	&	-1.55	&	 4055974360527366272	&	17.7	&	2.4	&	T	&	S	&		\\
LTNF 1	&	144.81	&	65.85	&	 786919754746647424	&	15.1	&	0.4	&	T	&	Bas	&	2	&	PN M  3-16	&	359.18	&	-2.30	&	 4056131006637615488	&	17.1	&	1.1	&	T	&	Em	&	2	\\
NGC 2371	&	189.16	&	19.84	&	 885587110718845568	&	14.8	&	-0.4	&	T	&	Bmps	&		&	PN PM 1-166	&	359.24	&	1.22	&	 4060159376692627840	&	15.3	&	3.2	&	P	&	B	&		\\
PN MaC 2-1	&	205.87	&	-26.73	&	 3211200438511961088	&	15.8	&	-0.4	&	T	&	S	&		&	PN M  3-44	&	359.39	&	-1.81	&	 4056355822397882880	&	16.0	&	2.1	&	T	&	B	&		\\
PN A66 30	&	208.56	&	33.29	&	 660071056749861888	&	14.4	&	-0.2	&	T	&	Ramrs	&	1	&	PN Th 3-35	&	359.39	&	1.40	&	 4060214180437611904	&	19.8	&	2.8	&	T	&	S	&	1	\\
PHR J0650+0013	&	212.64	&	-0.07	&	 3113542949606809088	&	15.2	&	1.1	&	T	&	Bmps	&		&	Terz N 19	&	359.89	&	5.25	&	 4109665712365779968	&	19.0	&	1.1	&	T	&	B	&		\\

\hline
\end{tabular}}
\begin{tablenotes}
      \tiny{
      \item (1) \citet{Jacoby21}; (2) \citet{Miszalski09}; (3) \citet{Hillwig16}; (4) \citet{Jones10}; (5) \citet{Pollacco94};(6) \citet{Afsar08}; (7) \citet{Corradi15}; \\
          (8) \citet{Douchin15};  (9) \citet{Miszalski11}; (10) \citet{Munday20}; (11) \citet{Aller20}; (12) \citet{Exter03}; (13) \citet{Ciardullo99}}.
    \end{tablenotes}
  \end{threeparttable}
\end{table*}

\begin{table*}
\tiny
\centering
\caption{The variable symbiotic stars in Gaia DR3.} \label{Table2}
\scalebox{1.1}{
\begin{tabular}{llllllll}
\hline
\#	&	target$\_$id	&	l	&	b	&	Gaia DR3 designation	&	G	&	B-R	&	Type	\\[3pt]
\hline
1	&	PN ShWi 5	&	1.21	&	-3.90	&	Gaia DR3 4050209822908746240	&	15.3	&	1.2	&	Symbiotic Star	\\
2	&	PN H 1-45	&	2.02	&	-2.06	&	Gaia DR3 4062646712567004416	&	14.3	&	3.0	&	Symbiotic Star	\\
3	&	PN Ap 1-11	&	3.12	&	-4.63	&	Gaia DR3 4050848540419995776	&	13.2	&	3.1	&	Symbiotic Star	\\
4	&	PN H 2-43	&	3.49	&	-4.87	&	Gaia DR3 4050670827750135040	&	13.6	&	1.2	&	Symbiotic Star	\\
5	&	IRAS 17554-2628	&	3.58	&	-1.22	&	Gaia DR3 4064034330564300928	&	19.8	&	2.8	&	Symbiotic Star	\\
6	&	PN M 3-18	&	7.57	&	1.44	&	Gaia DR3 4070389125449668608	&	11.9	&	5.0	&	Symbiotic Star	\\
7	&	PN Th 4-4	&	8.31	&	3.73	&	Gaia DR3 4119029875002043392	&	14.6	&	3.4	&	Symbiotic Star	\\
8	&	PN M  2-9	&	10.90	&	18.06	&	Gaia DR3 4335188603873318656	&	13.9	&	1.3	&	Symbiotic Star	\\
9	&	PN K  3-9	&	23.91	&	-1.54	&	Gaia DR3 4155672680486693120	&	15.3	&	2.6	&	Symbiotic Star	\\
10	&	PN Ap 3-1	&	37.64	&	-2.97	&	Gaia DR3 4268140453591785984	&	14.3	&	3.9	&	Symbiotic Star	\\
11	&	PN M  4-16	&	61.79	&	2.11	&	Gaia DR3 2022052808961769088	&	16.6	&	1.3	&	Symbiotic Star	\\
12	&	Hen 2-468	&	75.94	&	-4.44	&	Gaia DR3 1870194997404105856	&	12.6	&	2.8	&	Symbiotic Star	\\
13	&	PN M 1-2	&	133.12	&	-8.64	&	Gaia DR3 360112911622101120	&	12.6	&	1.2	&	Symbiotic Star	\\
14	&	Hen 2-34	&	274.19	&	2.58	&	Gaia DR3 5409069172514684416	&	14.7	&	2.7	&	Symbiotic Star	\\
15	&	Hen 2-25	&	275.22	&	-3.71	&	Gaia DR3 5310613021532357632	&	14.7	&	0.8	&	Symbiotic Star	\\
16	&	Hen 2-106	&	312.03	&	-2.03	&	Gaia DR3 5853777267581362176	&	13.3	&	0.8	&	Symbiotic Star	\\
17	&	Hen 2-104	&	315.48	&	9.46	&	Gaia DR3 6089564718596906880	&	13.6	&	0.8	&	Symbiotic Star	\\
18	&	Hen 2-134	&	319.22	&	-9.35	&	Gaia DR3 5822400362454690688	&	12.1	&	2.5	&	Symbiotic Star	\\
19	&	Hen 2-127	&	325.54	&	4.18	&	Gaia DR3 5889726659221998592	&	14.5	&	2.4	&	Symbiotic Star	\\
20	&	PN Cn 1-2	&	326.41	&	-10.94	&	Gaia DR3 5818044445302448000	&	10.6	&	1.8	&	Symbiotic Star	\\
21	&	PN Cn 1-1	&	330.78	&	4.15	&	Gaia DR3 5982979264021123968	&	10.8	&	1.1	&	Symbiotic Star	\\
22	&	Hen 2-156	&	338.94	&	5.36	&	Gaia DR3 5992529686406981248	&	12.4	&	2.1	&	Symbiotic Star	\\
23	&	Hen 2-176	&	339.39	&	0.74	&	Gaia DR3 5943382139466094720	&	13.6	&	4.1	&	Symbiotic Star	\\
24	&	Hen 2-171	&	346.03	&	8.55	&	Gaia DR3 6020686328090453888	&	14.9	&	5.5	&	Symbiotic Star	\\
25	&	PN H 2-4	&	352.95	&	3.93	&	Gaia DR3 5979902864926562176	&	14.2	&	3.2	&	Symbiotic Star	\\
26	&	PN M  2-24	&	356.99	&	-5.80	&	Gaia DR3 4042147516455759744	&	15.1	&	0.8	&	Symbiotic Star	\\
27	&	PN Th 3-20	&	357.41	&	2.62	&	Gaia DR3 4058701527427641472	&	14.0	&	2.8	&	Symbiotic Star	\\ \hline
	&		&		&		&		&		&		&		\\
28	&	PN K  4-26	&	37.18	&	-6.85	&	Gaia DR3 4263728319553777408	&	13.9	&	6.7	&	Mira Variable Candidate	\\
29	&	PN K  4-36	&	44.44	&	-10.38	&	Gaia DR3 4290522180961855872	&	12.7	&	4.7	&	M star	\\
30	&	CD-48 6027	&	283.90	&	9.73	&	Gaia DR3 5362804330246457344	&	12.1	&	-0.3	&	Hot Subdwarf	\\
31	&	Hen 2-58	&	289.18	&	-0.70	&	Gaia DR3 5338220285385672064	&	7.3	&	0.9	&	Wolf-Rayet	\\

\hline
\end{tabular}}
\end{table*}

\begin{table*}
\tiny
\centering
\caption{The CS radial velocity of PNe in Gaia DR3. The symbol (:) in column 4 refers to the RV measurement with high uncertainty.} \label{Table3}
\scalebox{1.00}{
\begin{tabular}{llccccccc}

\hline
\#    & PN name & \multicolumn{2}{c}{Galactic coordinate} && \multicolumn{2}{c}{RV (km/s)} & rv$\_$nb$\_$transits & rv$\_$visibility$\_$periods$\_$used \\ \cline{3-4} \cline{6-7}
      & 	                &	l	    &	 b	    &&	Gaia DR3	  &	\citet{Durand98}&       &      \\
\hline 			
1	&	MPA J1803-3043	    &	0.4	    &	-4.22	&&	              &	-130.5$\pm$6.0	&	3	&	2	\\
2	&	Ap 1-11	            &	3.12	&	-4.63	&&	62.0$\pm$3.9  &				    &	4	&	4	        \\
3	&	H 2-24	            &	4.33	&	1.84	&&	28.0$\pm$3.8  &	-198.2$\pm$	4.1	&	6	&	3	         \\
4	&	M 1-44	            &	4.97	&	-4.96	&&	-5.9$\pm$0.5  &	-75$\pm$11	    &	4	&	4	         \\
5	&	SB 15	            &	9.3	    &	-6.53	&&	-16.5	$\pm$	4.7	&	165	$\pm$	15	&	5	&	4	\\
6	&	PN G009.8-07.5	&	9.87	&	-7.56	&	&	-23.3	$\pm$	1.1	&	-32	$\pm$	30	&	2	&	2	\\
7	&	PN V-V 3-4	&	13.45	&	-4.25	&	&	-16.3	$\pm$	4.7	&				&	7	&	7	\\
8	&	UCAC4 374-117003	&	15.54	&	0.34	&	&	21.3	$\pm$	2.2	&				&	8	&	8	\\
9	&	SS 318	&	17.02	&	11.1	&	&	-36.7	$\pm$	2.4	&				&	19	&	11	\\
10	&	K 2-7	&	19.41	&	-19.66	&	&	-18.4	$\pm$	0.5	&				&	14	&	12	\\
11	&	PN G019.5-04.9	&	19.53	&	-4.96	&	&	-20.9	$\pm$	1.4	&				&	13	&	9	\\
12	&	Pe 1-15	&	25.91	&	-2.18	&	&	37.7	$\pm$	3.7	&				&	6	&	4	\\
13	&	IPHASX J191716.4+033447	&	39.08	&	-4.1	&	&	-4.2	$\pm$	8.6:	&				&	9	&	9	\\
14	&	K 1-14	&	45.61	&	24.32	&	&	-19.0	$\pm$	1.0	&				&	63	&	22	\\
15	&	VSP 2-30	&	49.32	&	2.38	&	&	-13.4	$\pm$	1.9	&				&	27	&	18	\\
16	&	Me 1-1	&	52.54	&	-2.96	&	&	-1.5	$\pm$	11.4:	&	-6	$\pm$	7	&	10	&	8	\\
17	&	NGC 7008	&	93.41	&	5.49	&	&	-71.9	$\pm$	7.7	&	-75.7	$\pm$	2.7	&	18	&	11	\\
18	&	IRAS 21282+5050	&	93.99	&	-0.12	&	&	28.2	$\pm$	7.3	&				&	20	&	15	\\
19	&	K 1-6	&	107.04	&	21.38	&	&	-55.7	$\pm$	13.7	&				&	13	&	13	\\
20	&	A 82	&	114.07	&	-4.67	&	&	-38.4	$\pm$	2.3	&	-30.5	$\pm$	3.3	&	23	&	16	\\
21	&	PN M 1-2	&	133.12	&	-8.63	&	&	-28.4	$\pm$	8.5	&	-12.1	$\pm$	2	&	13	&	11	\\
22	&	WeBo 1	&	135.67	&	1	&	&	-25.4	$\pm$	4.6	&				&	33	&	17	\\
23	&	NGC 1514	&	165.53	&	-15.29	&	&	42.3	$\pm$	1.3	&	59.8	$\pm$	4.4	&	21	&	10	\\
24	&	H 3-75	&	193.65	&	-9.58	&	&	11.6	$\pm$	2.6	&	22.9	$\pm$	2	&	26	&	11	\\
25	&	NGC 2346	&	215.7	&	3.62	&	&	31.5	$\pm$	3.0	&	21.8	$\pm$	0.9	&	13	&	10	\\
26	&	PHR j0701-0749	&	221	&	-1.41	&	&	44.9	$\pm$	2.9	&				&	14	&	11	\\
27	&	LoTr 1	&	228.21	&	-22.14	&	&	15.8	$\pm$	19.5:	&				&	31	&	19	\\
28	&	PN V-V 1-7	&	235.44	&	1.89	&	&	38.9	$\pm$	0.7	&				&	27	&	17	\\
29	&	WRAY 15-158	&	255.33	&	-3.64	&	&	24.4	$\pm$	2.3	&				&	22	&	19	\\
30	&	LoTr 3	&	265.11	&	-4.21	&	&	16.2	$\pm$	5.5	&	49	$\pm$	3	&	19	&	15	\\
31	&	NGC 3132	&	272.11	&	12.4	&	&	-11.1	$\pm$	1.6	&	-16	$\pm$	4.1	&	18	&	13	\\
32	&	Hen 2-36	&	279.61	&	-3.19	&	&	-0.1	$\pm$	1.5:	&	-7.1	$\pm$	2	&	18	&	10	\\
33	&	Hen 2-51	&	288.88	&	-5.22	&	&	13.5	$\pm$	3.9	&	8	$\pm$	3	&	13	&	12	\\
34	&	Al 1	&	291.1	&	-39.66	&	&	1.1	$\pm$	2.3:	&				&	16	&	15	\\
35	&	Hen 2-70	&	293.61	&	1.2	&	&	64.6	$\pm$	4.8	&				&	14	&	13	\\
36	&	PN A66 35	&	303.57	&	40	&	&	-38.1	$\pm$	4.6	&	-6.6	$\pm$	3.8	&	9	&	8	\\
37	&	SuWt 2	&	311.05	&	2.48	&	&	-15.7	$\pm$	7.5	&	-40	$\pm$	9	&	17	&	11	\\
38	&	MPA J1508-6455	&	316.77	&	-5.8	&	&	-39.2	$\pm$	11.9	&				&	43	&	17	\\
39	&	Hen 2-134	&	319.22	&	-9.35	&	&	9.0	$\pm$	0.7	&				&	45	&	21	\\
40	&	PM 1-89	&	324.09	&	3.53	&	&	-68.5	$\pm$	3.0	&	-81	$\pm$	20	&	13	&	12	\\
41	&	Hen 3-1312 (Sast 2-12)	&	334.84	&	-7.46	&	&	-70.6	$\pm$	1.0	&	-77	$\pm$	-7	&	25	&	13	\\
42	&	LoTr 5	&	339.89	&	88.46	&	&	-8.3	$\pm$	1.7	&				&	21	&	13	\\
43	&	 Vd 1-1	&	344.27	&	4.75	&	&	-70.6	$\pm$	1.0	&	-142.1	$\pm$	2.5	&	11	&	10	\\
44	&	SB 38	&	352.8	&	-8.41	&	&	-35.0	$\pm$	7.0	&	59	$\pm$	15	&	9	&	6	\\
45	&	PHR J1711-3210	&	353.28	&	4.25	&	&	27.2	$\pm$	1.7	&				&	10	&	6	\\
46	&	PN G354.8+01.6	&	354.89	&	1.63	&	&	13.6	$\pm$	6.6	&				&	13	&	10	\\
47	&	Pe 1-11	&	358.01	&	-5.16	&	&	-11.9	$\pm$	0.2	&	-130.6	$\pm$	14	&	12	&	5	\\
48	&	M 3-8	&	358.24	&	4.29	&	&	101.7	$\pm$	3.4	&	95	$\pm$	11	&		7   &	4\\
49	&	PHR J1752-3116	&	358.77	&	-2.5	&	&	-21.7	$\pm$	8.4	&				&	2	&	1	\\
50	&	Hen 3-1863	&	359.28	&	-33.5	&	&	4.3	$\pm$	10.2:	&				&	19	&	14	\\
51	&	Th 3-14	&	359.3	&	4.76	&	&	26.2	$\pm$	20.4:	&	-239.2	$\pm$	14	&	7	&	5	\\

\hline
\end{tabular}}
\end{table*}

\end{document}